\documentclass[review]{elsarticle}

\usepackage{lineno,hyperref}
\modulolinenumbers[250]
\journal{Journal of \LaTeX\ Templates}

\usepackage{hyperref}
\usepackage{latexsym}
\usepackage{float}
\usepackage{array}
\usepackage{rotating}
\usepackage{multirow}
\usepackage{amsmath,amssymb,graphicx} 
\usepackage{pstricks}
\usepackage{bbm}
\usepackage{latexsym}

\newcommand{\centeron}[2]{{\setbox0=\hbox{#1}\setbox1=\hbox{#2}\ifdim
\wd1>\wd0\kern.5\wd1\kern-.5\wd0\fi \copy0
\kern-.5\wd0\kern-.5\wd1\copy1\ifdim\wd0>\wd1
                                   \kern.5\wd0\kern-.5\wd1\fi}}
\newcommand{\ltap}{\>\centeron{\raise.35ex\hbox{$<$}}
                           {\lower.65ex\hbox{$\sim$}}\>}
\newcommand{\gtap}{\>\centeron{\raise.35ex\hbox{$>$}}
                           {\lower.65ex\hbox{$\sim$}}\>}

\newcommand\ZZ{\hbox{\zfont Z\kern-.4emZ}}
\font\zfont = cmss10 

\newcommand{\fref}[1]{Fig.~\ref{f.#1}}

\newcommand{\sref}[1]{Section~\ref{s.#1}}

\newcommand{\cref}[1]{Chapter~\ref{c.#1}}

\newcommand{\ba}{\begin{array}}
\newcommand{\ea}{\end{array}}

\newcommand{\beq}{\begin{eqnarray}}
\newcommand{\eeq}{\end{eqnarray}}
\newcommand{\beqs}{\begin{eqnarray*}}
\newcommand{\eeqs}{\end{eqnarray*}}

\newcommand{\bal}{\begin{align}} 
\newcommand{\eal}{\end{align}}

\def\bi{\begin{itemize}}
\def\ei{\end{itemize}}
\def\ben{\begin{enumerate}}
\def\een{\end{enumerate}}
\def\bc{\begin{center}}
\def\ec{\end{center}}
\def\bt{\begin{table}}
\def\et{\end{table}}
\def\btb{\begin{tabular}}
\def\etb{\end{tabular}}

\def\ev{\, {\rm  eV}}
\def\gev{\, {\rm GeV}}
\def\kev{\, {\rm keV}}
\def\mev{\, {\rm MeV}}
\def\tev{\, {\rm TeV}}

\def\mass2{mass${}^2$}

\def\cmmthree{\mathrm{cm}^{-3}}

\def\cmtwo{\mathrm{cm}^2}

\def\kmpers{\, \mathrm{km}/\mathrm{s}}

\usepackage{tikz}
\usetikzlibrary{arrows,shapes}
\usetikzlibrary{trees}
\usetikzlibrary{matrix,arrows} 				
\usetikzlibrary{positioning}				
\usetikzlibrary{calc,through}				
\usetikzlibrary{decorations.pathreplacing}  
\usepackage{pgffor}							

\usetikzlibrary{decorations.pathmorphing}	
\usetikzlibrary{decorations.markings}
\usetikzlibrary{snakes}

\tikzset{
    vector/.style={decorate, decoration={snake}, draw},
	provector/.style={decorate, decoration={snake,amplitude=2.5pt}, draw},
	antivector/.style={decorate, decoration={snake,amplitude=-2.5pt}, draw},
    fermion/.style={draw, postaction={decorate},
        decoration={markings,mark=at position .55 with {\arrow[draw]{>}}}},
    fermionbar/.style={draw, postaction={decorate},
        decoration={markings,mark=at position .55 with {\arrow[draw=black]{<}}}},
    fermionnoarrow/.style={draw},
    gluon/.style={decorate, draw,decoration={coil,amplitude=4pt, segment length=6pt}, line width=1},
    scalar/.style={dashed,draw, postaction={decorate},
        decoration={markings,mark=at position .55 with {\arrow[draw]{>}}}},
    scalarbar/.style={dashed,draw, postaction={decorate},
        decoration={markings,mark=at position .55 with {\arrow[draw]{<}}}},
    scalarnoarrow/.style={dash pattern = on 6 pt off 3 pt,draw},
    electron/.style={draw, postaction={decorate},
        decoration={markings,mark=at position .55 with {\arrow[draw]{>}}}},
	bigvector/.style={decorate, decoration={snake,amplitude=4pt}, draw},
	vectorscalar/.style={loosely dotted,draw, postaction={decorate}}
}

\begin{document} 
\begin{frontmatter}

\title{ \Large Direct Detection with Dark Mediators}

\author{David Curtin}\author{Ze'ev Surujon}
\address{C. N. Yang Institute for Theoretical Physics, Stony Brook University, 
 Stony Brook, NY 11794, U.S.A.}
\author{Yuhsin Tsai}
\address{Physics Department, University of California Davis, Davis, California 95616, U.S.A.}
  
\begin{abstract}
We introduce \emph{dark mediator Dark matter} (dmDM) where the dark and visible sectors are connected by at least one light mediator $\phi$ carrying the same dark charge that stabilizes DM. $\phi$ is coupled to the Standard Model via an operator $\bar q q \phi \phi^*/\Lambda$, and to dark matter via a Yukawa coupling $y_\chi \overline{\chi^c}\chi \phi$. 
Direct detection is realized as  the $2\rightarrow3$ process $\chi N \rightarrow \bar \chi N \phi$ at tree-level for $m_\phi \lesssim 10 \kev$ and small Yukawa coupling, or alternatively as a loop-induced $2\rightarrow2$ process $\chi N \rightarrow \chi N$. 
We explore the direct-detection consequences of this scenario and find that a heavy $\mathcal{O}(100 \gev)$ dmDM candidate fakes different $\mathcal{O}(10 \gev)$ standard WIMPs in different experiments. Large portions of the dmDM parameter space are detectable above the irreducible neutrino background and not yet excluded by any bounds. Interestingly, for the $m_\phi$ range leading to novel direct detection phenomenology, dmDM is also a form of Self-Interacting Dark Matter (SIDM), which resolves inconsistencies between dwarf galaxy observations and numerical simulations.
\end{abstract}


\end{frontmatter}

\linenumbers
\section{Introduction}
In this letter, we present \emph{Dark Mediator Dark Matter} (dmDM) to address two important gaps in the DM literature: exploring mediators with dark charge, and non-standard interaction topologies for scattering off nuclei. Additional details and constraints will be explored in a companion paper~\cite{dmDM}.

The existence of dark matter is firmly established by many astrophysical and cosmological observations~\cite{Ade:2013zuv}, but its mass and coupling to the Standard Model (SM) particles are still unknown. Weakly Interacting Massive Particles (WIMPs) are the most popular DM candidates since they arise in many theories beyond the SM, including supersymmetry, and may naturally give the correct relic abundance \cite{Jungman:1995df}. However, improved experimental constraints -- from collider searches, indirect detection and direct detection \cite{Goodman:1984dc, Akerib:2013tjd,Aprile:2012nq, Bernabei:2013cfa, Aalseth:2012if, Angloher:2011uu, Agnese:2013rvf,Fan:2013faa} -- begin to set tight limits (with some conflicting signal hints) on the standard WIMP scenario with a contact interaction to quarks. This makes it necessary to look for a more complete set of DM models which are theoretically motivated while giving unique experimental signatures.

\begin{figure}
\begin{center}
\includegraphics[width=8cm]{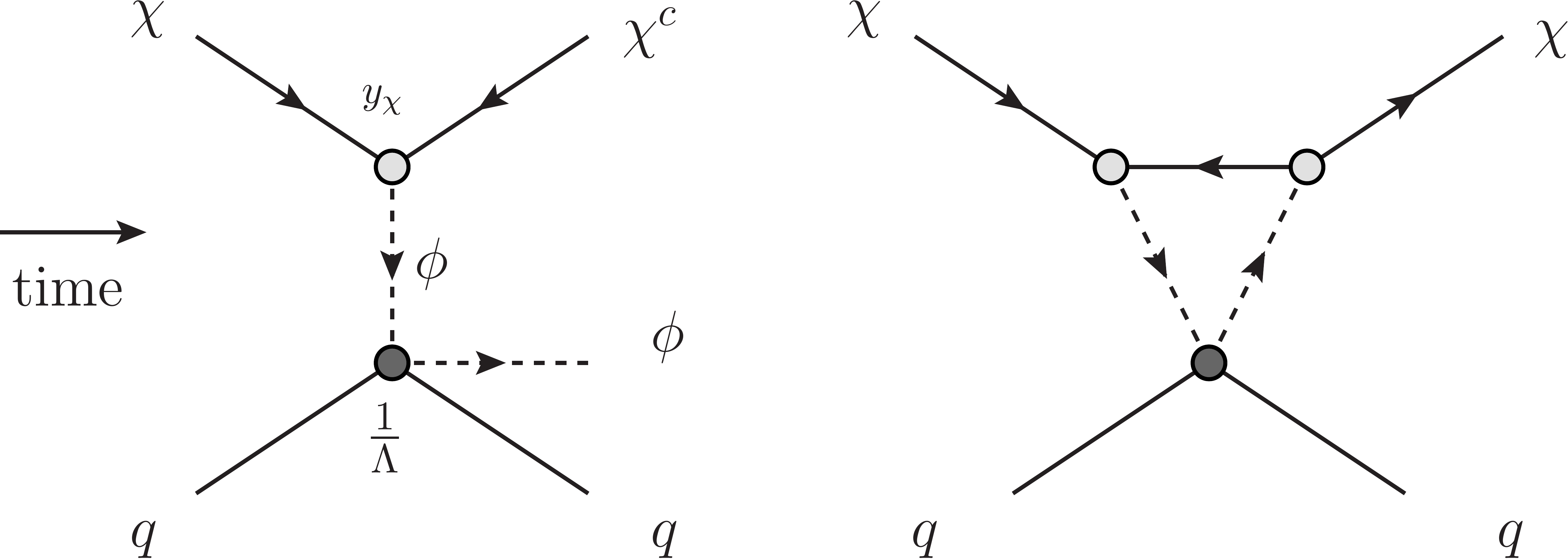}
\end{center}
\caption{
The quark-level Feynman diagrams responsible for DM-nucleus scattering in \emph{Dark Mediator Dark Matter} (dmDM). Left: the $2\rightarrow3$ process at tree-level. Right: the loop-induced $2\rightarrow2$ process. The arrows indicate flow of dark charge.
}
\label{f.feynmandiagram}
\end{figure}

\section{Dark Mediator Dark Matter}
Given its apparently long lifetime, most models of DM include some symmetry under which the DM candidate is charged to make it stable. An interesting possibility is that not only the DM candidate, but also the mediator connecting it to the visible sector is charged under this dark symmetry. Such a `dark mediator' 
$\phi$ could only couple to the SM fields in pairs.

As a simple example, consider real or complex SM singlet scalars $\phi_i$ coupled to quarks,
along with Yukawa couplings to a Dirac fermion DM candidate $\chi$.
The terms in the effective Lagrangian relevant for direct detection are
\begin{equation}
\mathcal{L}_\mathrm{DM}  \supset 
 \displaystyle{\sum_{i,j}^{n_{\phi}}}\,\frac{1}{\Lambda_{ij}} \bar q\,q \,\phi_i \phi_j^* + \displaystyle{\sum_{i}^{n_{\phi}}}\left ( y^{\phi_i}_\chi \overline{\chi^c}\chi \phi_i + h.c. \right)+...
\label{eq:dmDM}
\end{equation}
where $\ldots$ stands for $\phi, \chi$ mass terms, as well as the rest of the dark sector, which may be more complicated than this minimal setup.  This interaction structure can be enforced by a $\mathbb{Z}_4$ symmetry. To emphasize the new features of this model for direct detection, we focus on the minimal case with a single mediator $n_\phi = 1$ (omitting the $i$-index). However, the actual number of dark mediators is important for interpreting indirect constraints \cite{dmDM}.

The leading order process for DM-nucleus scattering is $\chi N \to \bar \chi N \phi$ if $m_\phi \lesssim \mathcal{O}(10 \kev)$. However, an elastic scattering $\chi N \to \chi N$ is always present at loop-level since it satisfies all possible symmetries, see \fref{feynmandiagram}. Which of the two possibilities dominates direct detection depends on the size of the Yukawa couplings $y_\chi^{\phi_i}$ as well as the dark mediator masses.

Previous modifications to WIMP-nucleon scattering kinematics include the introduction of a mass splitting  \cite{TuckerSmith:2001hy, Graham:2010ca, Essig:2010ye}; considering matrix elements $|\mathcal{M}|^2$ with additional velocity- or momentum transfer suppressions (for a complete list see e.g. \cite{MarchRussell:2012hi}), especially at low DM masses close to a GeV \cite{Chang:2009yt}; light scalar or `dark photon' mediators (see e.g. \cite{Essig:2013lka} which give large enhancements at low nuclear recoil); various forms of composite dark matter \cite{Alves:2009nf, Kribs:2009fy, Lisanti:2009am, Cline:2012bz, Feldstein:2009tr} which may introduce additional form factors; DM-nucleus scattering with intermediate bound states \cite{Bai:2009cd} which enhances scattering in a narrow range of DM velocities; and induced nucleon decay in Asymmetric Dark Matter models \cite{Davoudiasl:2011fj}.
Notably missing from this list are alternative process topologies for DM-nucleus scattering.
This omission is remedied by the dmDM scenario. 

dmDM is uniquely favored to produce a detectable $2\to3$ scattering signal at direct detection experiments. This is because it contains two important ingredients: (1) a light mediator with non-derivative couplings to enhance the cross section, compensating for the large suppression of emitting a relativistic particle in a non-relativistic scattering process, and (2) a scalar as opposed to a vector mediator, allowing it to carry dark charge (without a derivative coupling). This imposes selection rules which make the $2\to2$ process subleading in $y_\chi$. These ingredients are difficult to consistently implement in other model constructions without violating  constraints on light force carriers.

The effect of strong differences between proton and neutron coupling to DM have been explored by \cite{Feng:2011vu}. To concentrate on the kinematics we shall therefore assume the operator $\bar q q \phi \phi^*/\Lambda$ is flavor-blind in the quark mass basis. Above the electroweak symmetry breaking scale this operator is realized as $\bar Q_L H q_R \phi\phi^*/M^2$. It can be generated by integrating out heavy vector-like quarks which couple to the SM and $\phi$ \cite{dmDM}, giving  $1/\Lambda = y_Q^2 y_h v/M_Q^2$. This UV completion allows for large direct detection cross sections without being in conflict with collider bounds, but may be still probed at the 14 TeV LHC.

\begin{figure}
\begin{center}
\begin{tabular}{l}
 $\scriptstyle m_\chi = 10 \gev, m_\phi = 0.2 \kev, v = 400 \kmpers $
 \\
 $\scriptstyle m_N \ = \ \textcolor{blue}{28}, \  \textcolor{red}{73}, \ \textcolor{purple}{131} \gev$
 \end{tabular}
 \\
\includegraphics[width=7cm]{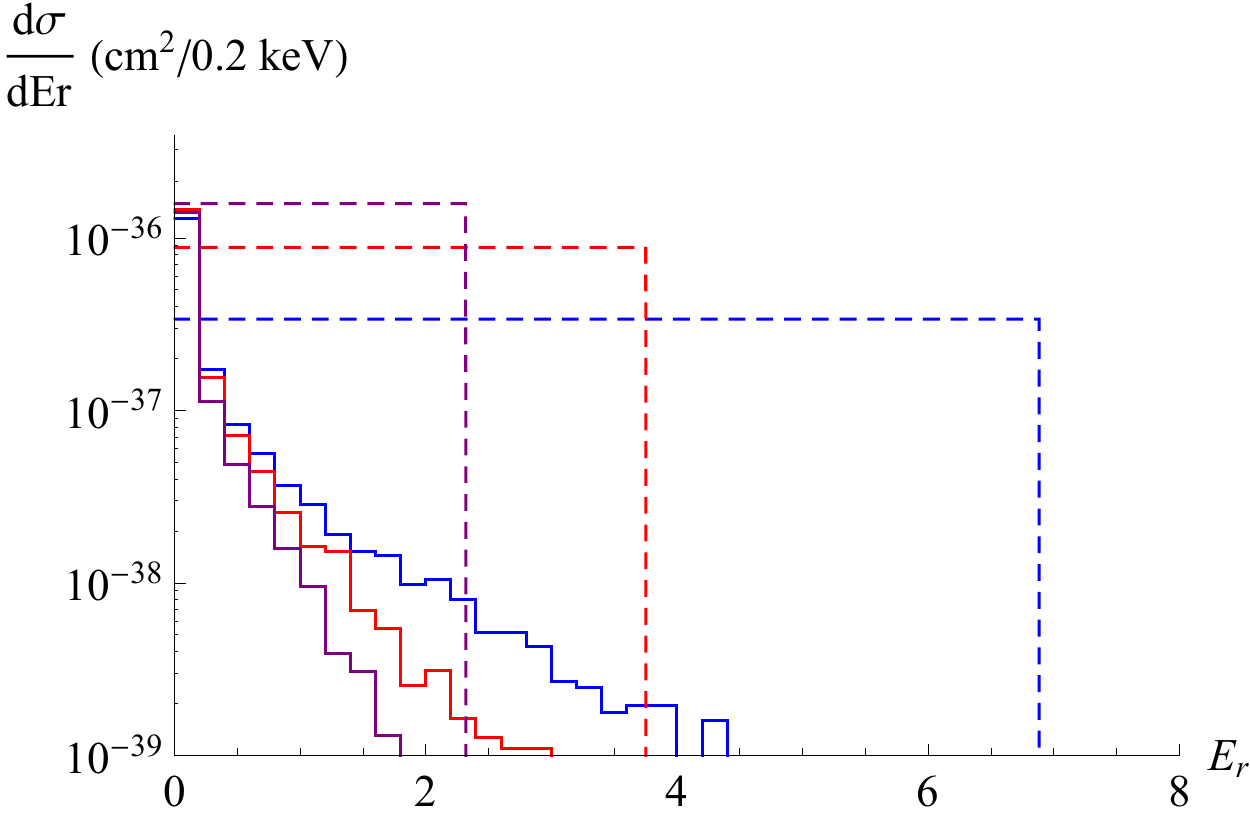}
\end{center}
\caption{
Nuclear recoil spectra of dmDM (without nuclear/nucleus form factors and coherent scattering enhancement) for  $y_\chi = 1, \Lambda = 1 \tev$ in a Silicon, Germanium and Xenon target. The dashed lines are spectra of standard WIMP scattering (via operator $\bar q q \bar \chi \chi/\tilde \Lambda^2$, with $\tilde \Lambda = 7 \tev$) shown for comparison. dmDM spectra computed with \texttt{MadGraph5} \cite{Alwall:2011uj} and  \texttt{FeynRules1.4} \cite{Ask:2012sm}. 
}
\label{f.partonleveldistribution}
\end{figure}

\section{Nuclear Recoil Spectrum} 
We start by examining the novel $2\rightarrow3$ regime of dmDM. The DM-nucleus collision is inelastic, not by introducing a new mass scale like a splitting, but by virtue of the process topology. The nuclear recoil spectrum is different compared to previously explored scenarios. This is illustrated in \fref{partonleveldistribution}, where we compare nuclear recoil spectra of standard WIMPs to dmDM for fixed velocity and different nucleus mass, \emph{before} convolving with various form factors and the ambient DM speed distribution. The observable dmDM differential cross section is independent of $m_\phi$ for $m_\phi \lesssim \kev$ and can be well described by the function
\begin{eqnarray}
\label{eq:recoil}
\frac{d\,\sigma_{2\to3}}{d\,E_r}&\simeq& \, \frac{\mathcal{C}}{E_r} \,\left(1-\sqrt{\frac{E_r}{E_r^\mathrm{max}}}\right)^2,
\end{eqnarray}
where $\mathcal{C}=1.3\times 10^{-42}\,(\tev/\Lambda)^2$ cm$^2$
and $E_r^\mathrm{max}=\frac{2 \mu_{\chi N}^2}{m_N} v^2 $, same as the WIMP case for a given DM velocity. (We emphasize that this is a phenomenological description, the actual spectra were produced in MadGraph, see \sref{directdetection}.)  The first factor comes from the light mediator propagator  $(2m_N\,E_r)^{-2}$ as well the integrated phase space of the escaping $\phi$. The cross section suppression (second factor) is more pronounced as the DM becomes lighter or slower, and as the nucleus becomes heavier, both of which reduces $E_r^\mathrm{max}$. This is because massless $\phi$ emission carries away a more significant fraction of the total collision energy if the heavy particle momenta are smaller. 
The maximum kinematically allowed nuclear recoil is then less likely.

When $n_{\phi}=1$, the $2\rightarrow2$ process will dominate direct detection for Yukawa coupling $y_\chi$ above some threshold, or if $m_\phi \gtrsim 10 \kev$. For the purpose of calculating the matrix element, the loop diagram in \fref{feynmandiagram} (right) is equivalent to the operator
\begin{equation}
\label{e.2to2operator}
\frac{\,y_{\chi}^2}{2\,\pi^2}\,\frac{1}{\Lambda\,q} \ (\bar{\chi}\,\chi\,\bar{N}\,N),
\end{equation}
where $q=\sqrt{2\,m_N\,E_r}$ is the momentum transfer in the scattering. Effectively, this is identical to a standard WIMP with a $\bar \chi \chi \bar N N$ contact operator, but with an additional $1/E_r$ suppression in the cross section. This gives a similar phenomenology as a light mediator being exchanged at tree-level with derivative coupling. 

Note that the relative importance of these two scattering processes is highly model dependent. For example, if $n_\phi = 2$ the dominant scalar-DM coupling could be $\bar q q \phi_1 \phi_2^*/\Lambda_{12}$. In that case, the $2\to2$ operator above is $\propto y_\chi^{\phi_1} y_\chi^{\phi_2}$ and can be suppressed without reducing the $2\to3$ rate by taking $y_\chi^{\phi_2} \gg y_\chi^{\phi_1}$. The scattering behavior of both the $2\rightarrow3$ and $2\rightarrow2$ regimes  necessitates a re-interpretation of all DM direct detection bounds. We will do this below.

\section{Indirect Constraints}
\label{s.indirectconstraints}

Direct detection experiments probe the ratio $y_\chi/\Lambda$ and $y_\chi^2/\Lambda$ for $2\to3$ and $2\to2$ scattering respectively. However, indirect constraints on dmDM from cosmology, stellar astrophysics and collider experiments are sensitive to the Yukawa coupling and $\Lambda$ separately. In \cite{dmDM} we conduct an extensive study of these bounds, including the first systematic  exploration of constraints on the $\bar q q \phi \phi^*/\Lambda$ operator with light scalars $\phi$. Since these constraints (in particular, Eqns.~\ref{e.NScoolingbound} and \ref{e.thermalrelic} below) provide important context for our results on direct detection, we summarize the two most important results here. For details we refer the reader to \cite{dmDM}.

The scalar mediator(s) of dmDM are most stringently constrained from stellar astrophysics and cosmology:
\begin{itemize}
\item
Avoiding overclosure requires $m_\phi \lesssim \ev$ \cite{Kolb:1990vq}, so we take the heaviest stable $\phi$ to be essentially massless, making it a very subdominant dark matter component. This also satisfies structure formation, computed for light sterile neutrinos in \cite{Wyman:2013lza}. Measurements by the Planck satellite \cite{Ade:2013zuv} restrict the number of light degrees of freedom during Big Bang Nucleosynthesis, enforcing the bound $n_\phi \leq 2$ for real scalars. 

\item
The coupling of $\phi$ to the SM is most constrained from stellar astrophysics.  For $n_\phi = 1$, observational data on neutron star cooling essentially rules out any directly detectable dmDM model~\cite{dmDM}.  However, this bound is easily relaxed for $n_\phi = 2$ if $m_{\phi_1} \lesssim \ev$, $m_{\phi_2} \sim \mev$, with a cosmologically unstable $\phi_2$. The dominant interaction to the SM is assumed to be $\bar q q \phi_1 \phi_2^*/\Lambda$. In that case, $\phi_2$ emission in the neutron star is Boltzmann suppressed due to its core temperature of $T \lesssim 100 \kev$, and $\phi_1$ emission proceeds via a loop process. The bound on $\Lambda$ is then weakened to 
\begin{equation}
\label{e.NScoolingbound}
\Lambda \gtrsim 10 \tev.
\end{equation}

\item
In Supernovae, emission of light invisible particles  can truncate the neutrino burst \cite{Raffelt:2006cw}.  However, if these particles interact with the stellar medium more strongly than neutrinos they are trapped and do not leak away energy from the explosion. The temperature of supernovae $T \sim 10 \mev$ is large enough to produce $\phi_1, \phi_2$ at tree-level in the above $n_\phi = 2$ scenario, and the scattering cross section with nuclei is much larger than for neutrinos if $\Lambda \lesssim 10^6 \tev$. Therefore this setup is compatible with supernovae constraints. 

\item
The LHC can set constraints on heavy dark vector quarks in a possible UV completion of dmDM. The CMS $20 \ \mathrm{fb}^{-1}$ di-jet + MET search \cite{CMS:2013gea} search sets a lower bound on the heavy quark to be $1.5$ TeV. 
\end{itemize}
The physics of direct detection for this $n_\phi = 2$ setup is identical to the minimal $n_\phi = 1$ model. This is because the typical momentum transfer is $\mathcal{O}(10 \mev)$, making the intermediate $\phi_2$ mediator massless for the purposes of direct detection. We are therefore justified in examining the direct detection phenomenology of the $n_\phi = 1$ model in detail, applying the $\Lambda$ bound Eqn.~\ref{e.NScoolingbound} and with the understanding that the full realization of dmDM requires a slightly non-minimal spectrum. 

The dark matter yukawa coupling is constrained from observations on large scale structure and (under certain assumptions) from cosmology:
\begin{itemize}
\item
Dark matter self-interaction bounds from bullet cluster observations constrain the DM Yukawa coupling to be $y_\chi \lesssim 0.13 (m_\chi/\gev)^{3/4}$ \cite{Feng:2009mn}.
\item A thermal relic $\chi$ with $\Omega_\chi = \Omega_\mathrm{CDM}$ requires 
\begin{equation}
\label{e.thermalrelic}
y_\chi = y_\chi^\mathrm{relic}(m_\chi) \approx 0.0027  \left( \frac{m_\chi}{\gev}\right)^{1/2}
\end{equation}
if there is no significant $\phi^3$ term.  This also satisfies the above self-interaction bounds.
\end{itemize}
Interestingly, the range of $m_\phi \sim \ev$ to MeV that is relevant for its novel direct detection signal also makes dmDM a realization of Self-Interacting DM (SIDM) \cite{ Carlson:1992fn, Spergel:1999mh, Wandelt:2000ad, Loeb:2010gj,  Rocha:2012jg, Zavala:2012us,  Medvedev:2013vsa,  Tulin:2013teo}. A Yukawa interaction consistent with $\chi$ being a thermal relic can then help resolve the ``core/cusp'' and ``too-big-to-fail'' inconsistencies between dwarf galaxy observations and many-body simulations \cite{dmDM, Tulin:2013teo}.

\section{Direct Detection}
\label{s.directdetection}
We compute dmDM nuclear recoil spectra at direct detection experiments by simulating the parton-level process in  \texttt{MadGraph5} \cite{Alwall:2011uj}, and derive the event rates according to
\begin{equation}
\frac{d R}{d E_r} = N_T \frac{\rho_\chi}{m_\chi} \int dv \ v f(v) \frac{d \sigma_N}{d E_r},
\end{equation}
where
$f(v)$ is the local DM speed distribution (approximate Maxwell-Bolzmann with $v_0 \approx 220$ km/s and a $v_{esc} \approx 544$ km/s cutoff,  boosted into the earth frame $v_e \approx 233$ \cite{Smith:2006ym}), while $\rho_\chi \approx 0.3 \  \gev \  \cmmthree$ is the local DM density \cite{Bovy:2012tw}, and $N_T$ is the target number density per kg.  $d \sigma_N/d E_r$ includes the usual Helm nuclear form factor  \cite{Engel:1991wq, Lewin:1995rx}, the $A^2$ coherent scattering enhancement as well as the quark-nucleon form factor for scalar interactions (see \cite{Belanger:2008sj} for a review). We validated our Monte Carlo pipeline by reproducing analytically known $2\rightarrow2$ results.

\fref{DDspectraCDMS} shows some nuclear recoil spectra for a Silicon and Xenon target. (We henceforth assume an effectively massless $\phi$.) dmDM is compared to standard WIMPs (velocity- and recoil-independent contact-interaction) for different DM masses. An important  feature of our model is apparent: a $\sim 50 \gev$ dmDM candidate looks like a $\sim 10 \gev$ $(20 \gev$) WIMP when scattering off Silicon (Xenon). Moreover, the shape of $d \sigma_N/d E_r$ is insensitive to $m_\chi$ unless $m_\chi$ is much smaller than $m_N$ (see Eq.~\ref{eq:recoil}). This makes it much more difficult to measure the DM mass using the shape of the spectrum. Signals at two detectors with different target materials are required.

\begin{figure*}
\begin{center}
\includegraphics[width=15cm]{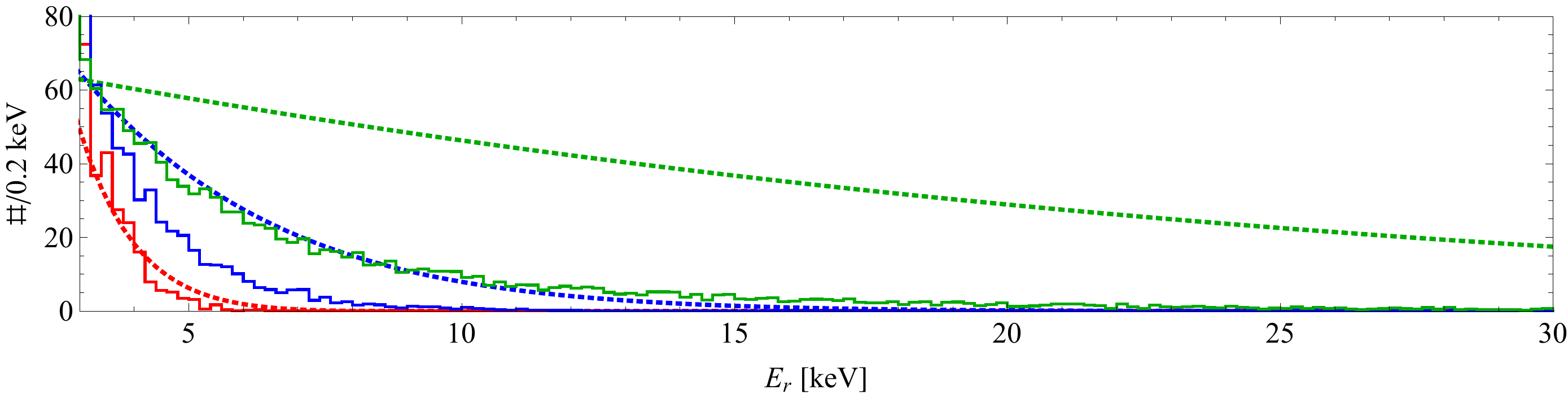}
\\
\includegraphics[width=15cm]{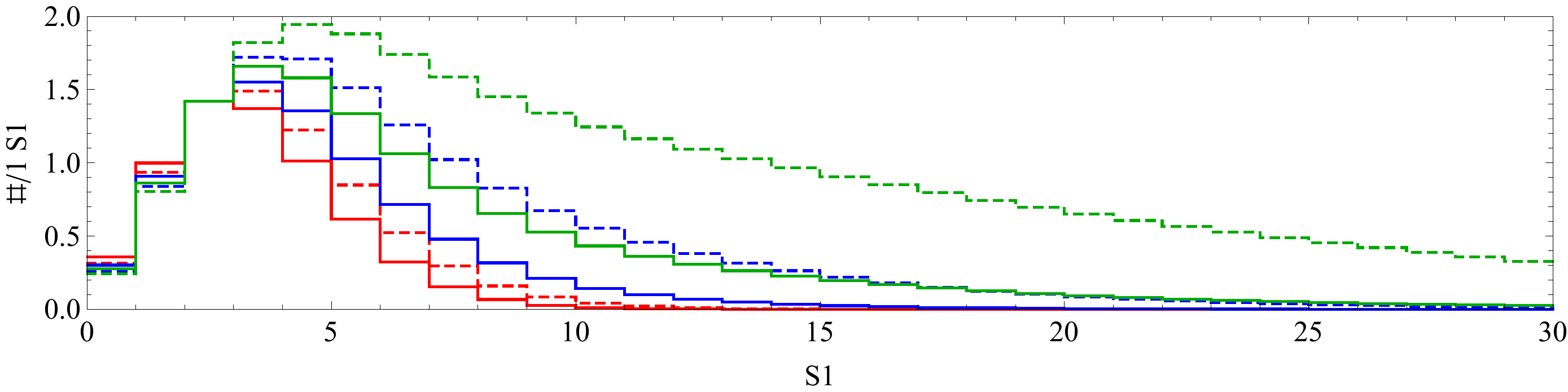}
\end{center}\vspace{-3mm}
\caption{
\textbf{Top:} Nuclear recoil spectra at CDMS II Silicon ($m_N = 28 \gev$) with 140.2 kg$\cdot$days exposure for dmDM (solid) and WIMP DM (dotted) of mass 5 (red), 10 (blue) and 50 (green) GeV. Experimental efficiencies are not included, and the recoil spectrum is shown only for $E_r > 3 \kev$ because the dmDM spectrum is so sharply peaked at the origin that no other features would be visible if it were included. The shown WIMP-nucleon cross sections for (5, 10, 50) GeV are $(4, 2, 6) \times 10^{-40} \ \cmtwo$, while the dmDM parameters are $y_\chi = 0.02$, $\Lambda = (29, 91, 91) \tev$ and $m_\phi < \kev$.
\textbf{Bottom:}
S1 spectra at LUX ($m_N = 131 \gev$) with 10065.4 kg$\cdot$days exposure for dmDM (solid) and WIMP DM (dotted) of mass 10 (red), 20 (blue) and 50 (green) GeV. The 14\% S1 light gathering efficiency is included but selection cuts are not. No DM signal below $E_r = 3 \kev$ is included due to limitations of the measured $\mathcal{L}_{eff}$, in accordance with the collaboration's analysis. The shown WIMP-nucleon cross sections for (10, 20, 50) GeV are $(18.5, 3.6, 4.9) \times 10^{-45} \ \cmtwo$, while the dmDM parameters are $y_\chi = 0.02$ and $\Lambda = (1900, 9700, 13000) \tev$ and $m_\phi < \kev$.
}
\label{f.DDspectraCDMS}
\end{figure*}

\begin{figure}
\begin{center}
\hspace{1.5mm}
\begin{tabular}{m{6.3cm}m{6.3cm}}
\includegraphics[width=6.3cm]{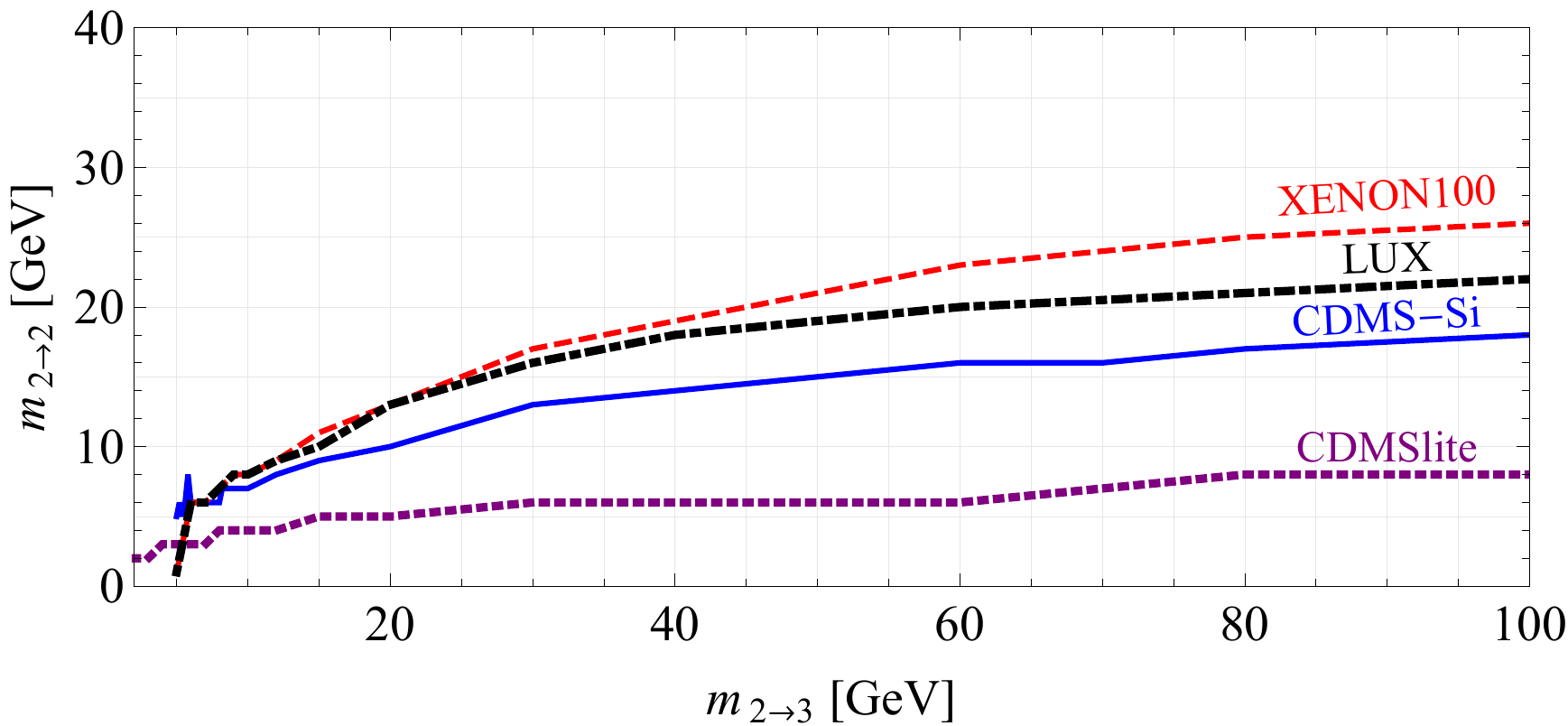}
&
\includegraphics[width=6.3cm]{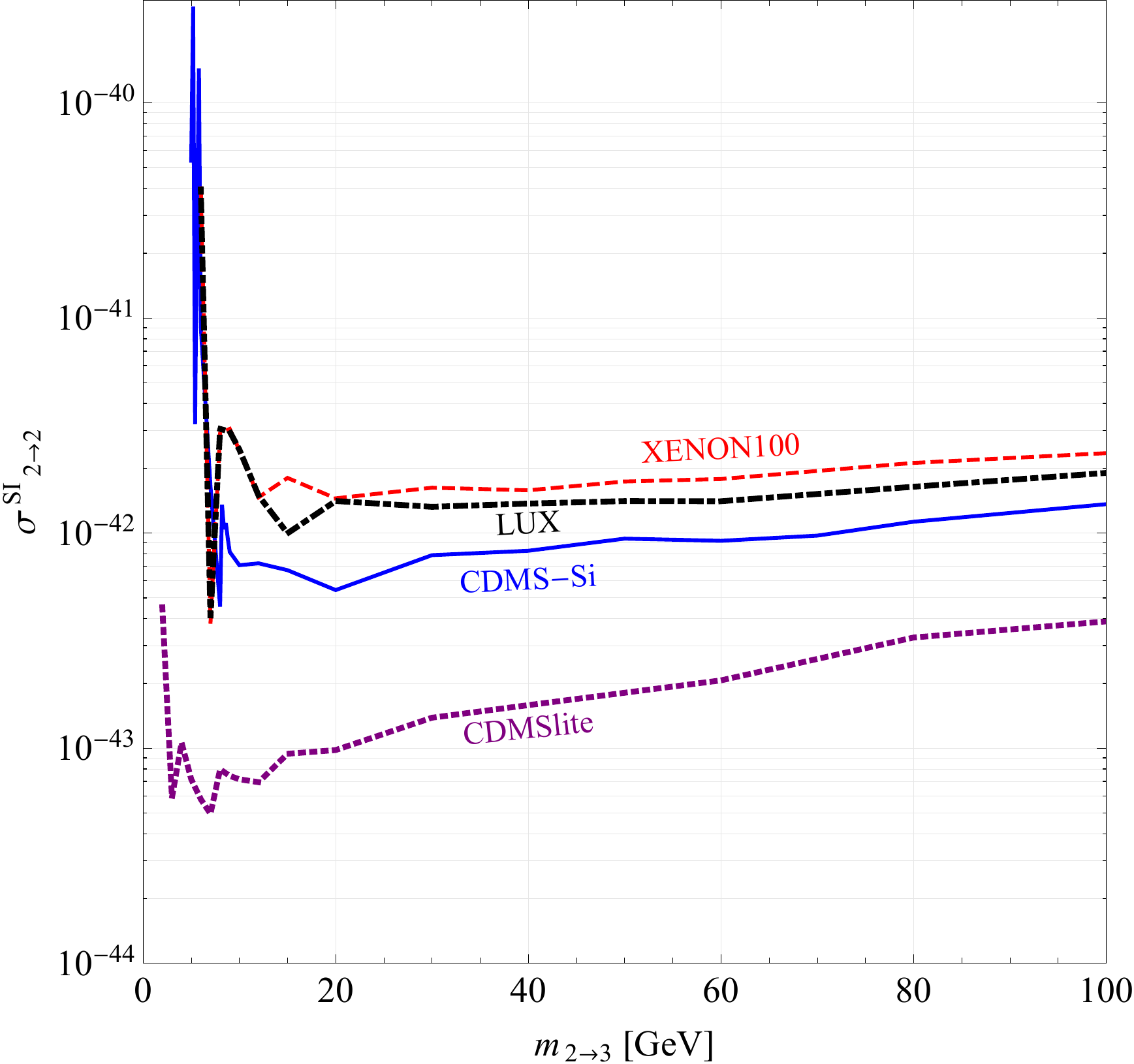}
\end{tabular}
\end{center}\vspace*{-2mm}
\caption{
\textbf{Left:} For each dmDM mass $m_\chi = m_{2\rightarrow3}$ this plot shows the WIMP mass $m_\chi = m_{2\rightarrow2}$ which gives the same spectral shape at XENON100 ($S1 > 3$ with 6\% light gathering efficiency, dashed red line), LUX ($S1 > 2$ with 14\% light gathering efficiency, dash-dotted black line), CDMSII Silicon ($E_r > 7 \kev$, solid blue line), and CDMSlite (Germanium, $Er > 0.2 \kev$, dotted purple line) before selection cuts.
\textbf{Right:} The `observed' WIMP-nucleon cross section for each dmDM mass $m_{2\to3}$, assuming the best-fit $m_{2\to2}$ from the left. The dmDM parameters are $y_\chi = 1, \Lambda = 45 \tev$.
}
\label{f.compare2to2to2to3}
\end{figure}



We can make this observation more concrete by mapping dmDM parameters to WIMP parameters.  This is possible because both sets of nuclear recoil spectra  look roughly like falling exponentials. For each dmDM spectrum  with a given mass there is a closely matching WIMP spectrum with some different (lower) mass.  To find the $m_{2\rightarrow2}$ corresponding to each $m_{2\rightarrow3}$ we compare binned WIMP and dmDM distributions and minimize the total relative difference in each bin. The resulting mapping is shown in \fref{compare2to2to2to3} (left).  Even very heavy dmDM candidates mimic light WIMPs of different masses at different experiments. A corresponding cross section remapping (right) shows that experiments with heavier nuclei are more sensitive to dmDM due to the inelastic nature of the collision.


\begin{figure}
\begin{center}
\hspace*{-7mm}
\includegraphics[width=8cm]{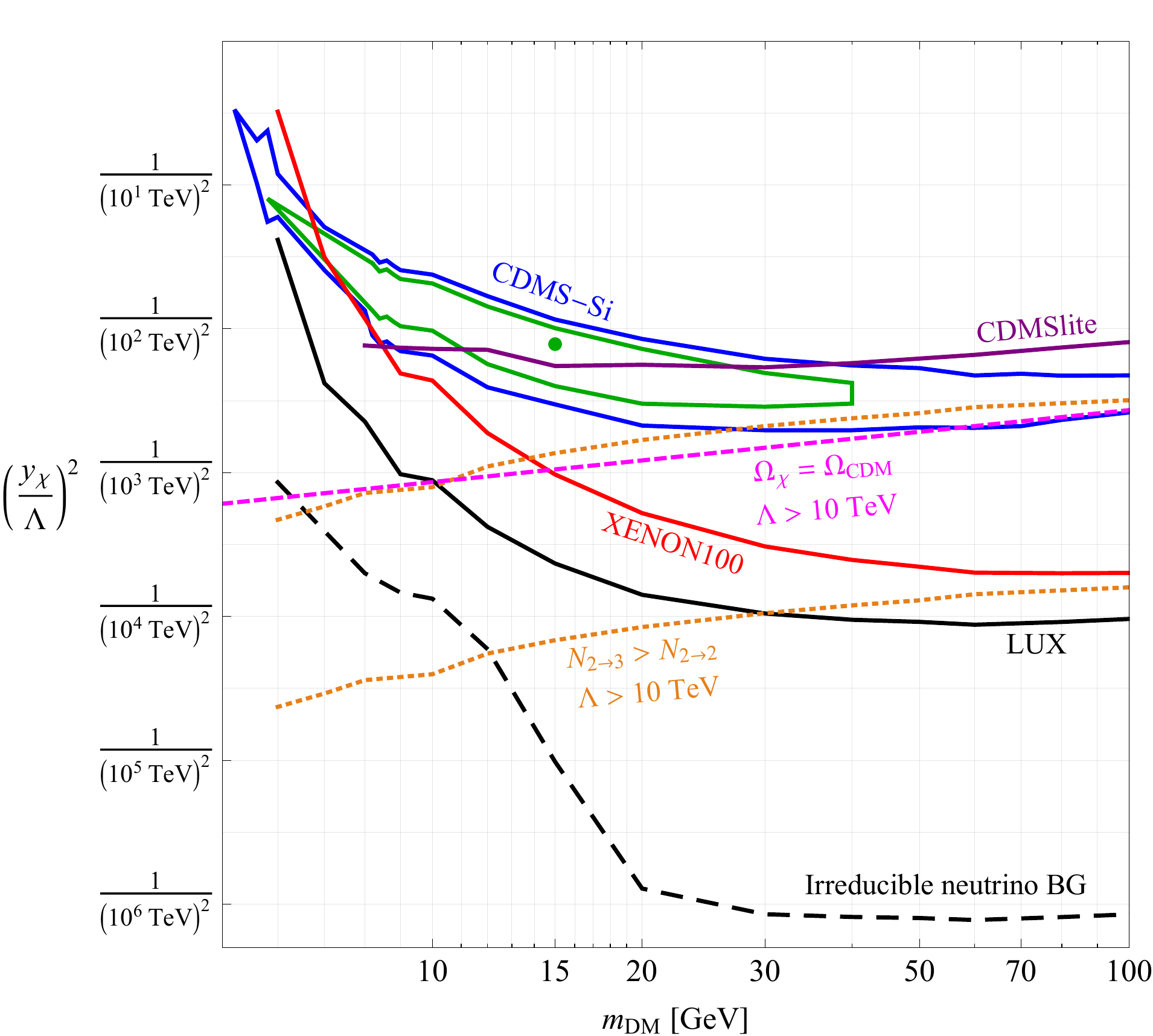} \includegraphics[width=8cm]{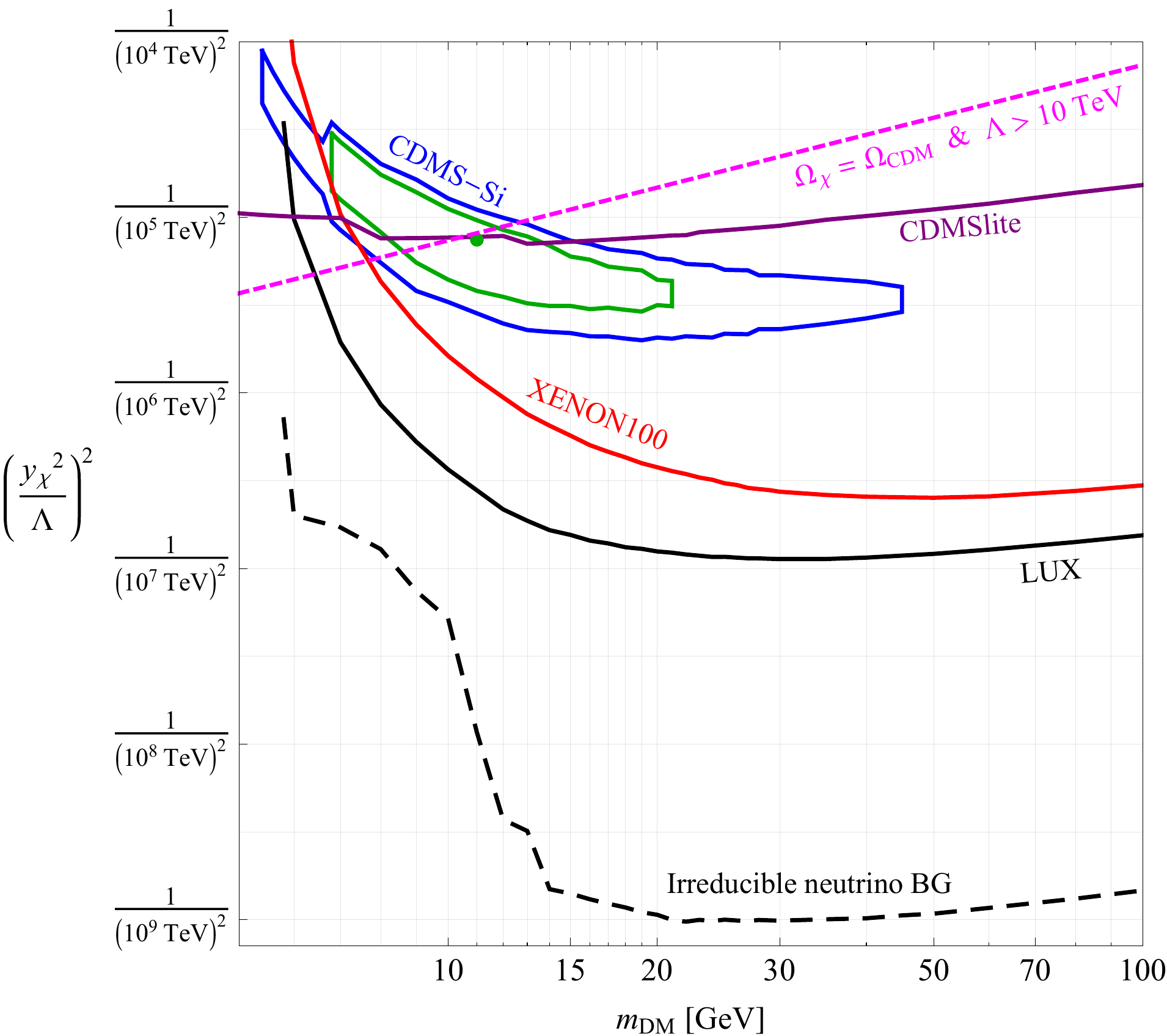}
\end{center}
\caption{\textbf{Left:} Direct detection bounds on the $2\rightarrow3$ regime of dmDM. The vertical axis is proportional to $\sigma_{\chi N \to \bar \chi N \phi}$. \emph{Solid lines}: 90\% CL bounds by XENON100 (red), LUX (black) and CDMSlite (purple), as well as the best-fit regions by CDMS II Si (blue, green). The large-dashed black line indicates the irreducible neutrino background \cite{Billard:2013qya}. 
\emph{Small-dashed magenta line}: Upper bound for $y_\chi = y_\chi^\mathrm{relic}(m_\chi)$ and neutron star cooling bound $\Lambda < 10 \tev$.
\emph{Lower dotted orange line}: upper bound for $2\to3$ dominated direct detection and neutron star bound with all equal Yukawa couplings. This line can be arbitrarily moved, as discussed below Eq.~\ref{e.2to2operator}. The \emph{upper dotted orange line} is for $y^{\phi_1}_\chi = y^{\phi_2}_\chi/20$, in which case the vertical axis is understood to be $(y_\chi^{\phi_2}/\Lambda)^2$.
\textbf{Right:} 
Direct detection bounds on the $2\rightarrow2$ regime of $n_\phi = 1$ dmDM, same labeling as the left plot. The vertical axis is proportional to $\sigma_{\chi N \to \bar \chi N}$, and is understood to be $(y_\chi^1 y_\chi^2/\Lambda)^2$ for the $n_\phi = 2$ model outlined in \sref{indirectconstraints}. 
}
\label{f.mappingbounds}
\end{figure}


\fref{compare2to2to2to3} defines an experiment-dependent parameter map that we can use to map each collaboration's WIMP bounds onto the dmDM model if $2\to3$ scattering dominates\footnote{We have confirmed the validity of this approach with full maximum likelihood fits \cite{Barlow:1990vc}.}. The resulting direct detection bounds are shown in \fref{mappingbounds} (left). We include the irreducible neutrino background \cite{Billard:2013qya} at the LUX experiment, which serves as an approximate lower border of the observable dmDM parameter space. 
An identical procedure can be used in the $2\to2$ dominant regime of dmDM. The translation map has similar qualitative features to the previous case since $d \sigma/d E_r \sim E_r^{-1}$, except the faked WIMP signal corresponds to somewhat higher mass. The resulting direct detection bounds are shown in \fref{mappingbounds} (right).

The probability for any one $2\rightarrow2$ nuclear recoil event to lie above experimental detection threshold is much larger than for a $2\rightarrow3$ event, due to the less severe recoil suppression. For $n_\phi = 1$, this means the former will dominate direct detection unless $m_\phi \lesssim \kev$ and the Yukawa coupling is very small, $y_\chi \lesssim 10^{-3} < y_\chi^\mathrm{relic}$. However, as discussed in \sref{indirectconstraints}, the neutron star cooling constraint requires at least $n_\phi = 2$. The $2\to2$ process could then be arbitrarily suppressed, allowing $2\to3$ direct detection with a thermal relic $\chi$.

For the $2\to3$ and $2\to2$ scattering regimes, direct detection probes $y_\chi/\Lambda$ and $y_\chi^2/\Lambda$ respectively. The neutron star cooling bound $\Lambda \gtrsim 10 \tev$ and the bounds on dark matter Yukawa coupling $y_\chi$ can be combined to be shown in the direct detection planes of \fref{mappingbounds}. The assumption of a thermal relic then excludes the regions in Fig. \ref{f.mappingbounds} above the magenta dashed line, meaning these bounds supersede the liquid Xenon experiments for $m_\chi \lesssim 10 \gev$ in the $2\to3$ dominant regime.

There are large discoverable regions of dmDM parameter space that are not excluded. Due to the nontrivial dependence of the dmDM recoil spectrum on the target- and dark-matter masses and velocity, signals at several experiments will be needed to differentiate standard WIMPs from our model, but dmDM offers the realistic prospect of TeV-scale heavy quark discoveries pointing the way towards a sensitivity target for direct detection.

\section{Conclusion} 
\emph{Dark Mediator Dark Matter} introduces the possibility that dark matter interacts with the standard model via a mediator which also carries dark charge.
This ``Double-Dark Portal'' adds the phenomenon of additional particle emission to the menu of possible interactions with nuclei, serving as an existence proof that this scattering topology can be realized. 
 Direct detection experiments are starting to probe interesting regions of parameter space compatible with a thermal relic and neutron star bounds. 
For observationally relevant parameters, dmDM also acts as an implementation of SIDM \cite{ Carlson:1992fn, Spergel:1999mh, Wandelt:2000ad, Loeb:2010gj,  Rocha:2012jg, Zavala:2012us,  Medvedev:2013vsa,  Tulin:2013teo}, which can resolve various inconsistencies between many-body simulations and observations for dwarf galaxies. 
Even more than many other DM models, dmDM discovery is aided by lowering nuclear recoil thresholds.  Further investigation is warranted and includes potential LHC signals, as well as possible leptophilic realizations of the model.

\textbf{Acknowledgements ---}
The authors would like to gratefully acknowledge the valuable contributions of Yue Zhao during an earlier stage of this collaboration. 
We thank 
Rouven Essig, 
Patrick Fox, 
Roni Harnik and 
Patrick Meade 
for valuable comments on draft versions of this letter. We are very grateful to 
Joseph Bramante,
Rouven Essig, 
Greg Gabadadze, 
Jasper Hasenkamp, 
Matthew McCullough,
Olivier Mattelaer,
Matthew Reece, 
Philip Schuster, 
Natalia Toro, 
Sean Tulin,
Neal Weiner, 
Itay Yavin and
Hai-Bo Yu
for valuable discussions. D.C. and Z.S. are supported in part by the National Science Foundation under Grant PHY-0969739. Y.T. is supported in part by the Department of Energy under Grant DE-FG02-91ER40674. The work of Y.T. was also supported in part by the National Science Foundation under Grant No. PHYS-1066293 and the hospitality of the Aspen Center for Physics

\end{document}